\documentclass[10pt,dvips,a4paper]{article}
\usepackage{epsfig}

\newcommand{\sbf}[1]{\boldsymbol{#1}}
\newcommand{\mbf}[1]{\mathbf{#1}}
\newcommand{\pard}[2]{\dfrac{\partial{#1}}{\partial{#2}}}
\newcommand{\veps}{\varepsilon}
\newcommand{\ignore}[1]{}

\newcommand*\patchAmsMathEnvironmentForLineno[1]{%
  \expandafter\let\csname old#1\expandafter\endcsname\csname #1\endcsname
  \expandafter\let\csname oldend#1\expandafter\endcsname\csname end#1\endcsname
  \renewenvironment{#1}%
     {\linenomath\csname old#1\endcsname}%
     {\csname oldend#1\endcsname\endlinenomath}}%
\newcommand*\patchBothAmsMathEnvironmentsForLineno[1]{%
  \patchAmsMathEnvironmentForLineno{#1}%
  \patchAmsMathEnvironmentForLineno{#1*}}%
\AtBeginDocument{%
\patchBothAmsMathEnvironmentsForLineno{equation}%
\patchBothAmsMathEnvironmentsForLineno{align}%
\patchBothAmsMathEnvironmentsForLineno{flalign}%
\patchBothAmsMathEnvironmentsForLineno{alignat}%
\patchBothAmsMathEnvironmentsForLineno{gather}%
\patchBothAmsMathEnvironmentsForLineno{multline}%
}

\usepackage{amssymb}
\usepackage{amsmath}
\usepackage{setspace}
\usepackage{lineno}

\usepackage[numbers]{natbib}

\begin{document}

\renewcommand{\title}{On a damage-plasticity approach to model concrete failure}

\begin{center} \begin{LARGE} \textbf{\title} \end{LARGE} \end{center}

\begin{center} 
Peter Grassl\\
Department of Civil Engineering\\
University of Glasgow, Glasgow, United Kingdom\\
Email: grassl@civil.gla.ac.uk, Phone: +44 141 330 5208, Fax: +44 141 330 4557
\end{center}

\section*{Abstract}
A damage-plasticity constitutive model for the description of fracture in plain concrete is presented.
Two approaches, the local model comprising the adjustment of the softening modulus and the nonlocal model based on spatial averaging of history variables, are applied to the analysis of a concrete bar subjected to uniaxial tension and to a three-point bending test.
The influence of mesh size and the decomposition into damage and plasticity components are discussed.
It is shown that for the two examples studied, both approaches result in mesh-independent results.
However, the nonlocal model, which relies on spatial averaging of history variables, exhibits sensitivity with respect to boundary conditions, which requires further studies. 

Keywords: Plasticity, damage mechanics, fracture, nonlocal, concrete

Revised version, Resubmitted to Engineering and Computational Mechanics, 2nd of January 2009

\section{Introduction}
Concrete is a strongly heterogeneous and random material, which in the hardened state consist mainly of aggregates, cement paste, air voids and water.
The heterogeneities range from several nano-meters in the structure of the cement paste to a couple of centimeters for aggregates.
Macroscopic fracture in this heterogeneous material is characterised by a quasi-brittle response.
Displacement controlled tests of the mechanical response of concrete in uniaxial tension, for instance, exhibit softening, which is defined in this context as a gradual decrease of the load with increasing displacements.
This softening response is caused by the development of a fracture process zone, involving mechanisms such as branching and bridging of micro-cracks.
The fracture process zone evolves from distributed and disconnected micro-cracks to a strongly localised tortuous macroscopic crack, whereby the tortuosity is governed by the presence of aggregates, which represent the dominant heterogeneity of the material. 

For the analysis of concrete structures, it is important to include the quasi-brittle softening response of concrete, since it strongly influences the load capacity of structures and is the source of a particular size effect on nominal strength, which neither follows the strength theory nor linear elastic fracture mechanics, see \cite{Baz01}.
However, the heterogeneities in concrete described above, are usually significantly smaller than the size of structures analysed, which makes it often not practical to model the heterogeneities explicitly.
Instead, concrete is usually assumed to be homogeneous on the scale studied and the formation of cracks is assumed to be an averaged process. 
For many heterogeneous materials, the averaged response can be obtained from spatial averaging of the material response on lower scales. 
However, for localised fracture, as in the case of concrete, spatial averaging alone is statistically insufficient, since energy is dissipated in localised zones.
Thus, additional to spatial averaging, averaging of many individual fracture process zones has to be performed.
This average of tortuous crack patterns results in a spatially distributed fracture process zone, which is governed by the size of the maximum heterogeneity. 
The fracture process zone is defined as the average of the zone in which energy is dissipated during the fracture process.
At a late stage of the fracture process, the fracture process zone in a single random analysis is fully localised in a single crack.
The deterministic nonlocal model represents the average of many random analyses.
Therefore, the average of fracture process zones at this stage of cracking is not localised but distributed, with its size determined by the tortuosity of crack paths.

Linear elastic fracture mechanics (LEFM) assumes a large stress-free crack with all the nonlinearities of the fracture process concentrated in an infinitesimally small zone in front of the crack tip \cite{Gri21}.
The basic aim of LEFM is to find the amount of energy available for crack growth and to compare it to the energy required to extend the crack \cite{Irw58, Ric68a, Ric68}. 
The assumption of an infinitesimal small fracture process zone allows for the combination of LEFM with the framework of linear elastic mechanics of materials. 
Therefore, LEFM has been used for decades successfully for the prediction of the failure of structures.
However, if the size of the fracture process zone is large compared to the length of the stress free crack and the size of the structure, LEFM results in a poor approximation of the fracture process. Most concrete structures are too small to fall in the range which is applicable for LEFM and a more accurate description by means of nonlinear fracture mechanics (NLFM) is required.
NLFM describes the nonlinear response within the fracture process zone by means of softening stress-strain relationships.

Constitutive models can be divided into local and nonlocal models. For local constitutive models, the stress at a point depends on the history of this point only.  
A local constitutive model for fracture of concrete represents an average of stochastic fracture process zones by means of a softening stress-strain curve.
Cracks are represented by localised strain profiles within finite elements, whereby the integral of the energy dissipation density over the size of the finite element results in the dissipated energy. 
Thus, for local softening constitutive models for fracture, the dissipated energy depends on the size of the finite element.
However, the energy dissipated during the fracture process of concrete is a material property and should be independent of the discretisation applied.
This limitation of local constitutive models can be overcome by adjustment of the softening modulus of the softening branch of the stress-strain curve with respect to the finite element size \cite{Pietruszczak81,BazOh83,WilBicStu86}.
With this approach, the dissipated energy is modelled mesh independently, as long as the inelastic strains localise in a zone of assumed size.
For more complex loading cases, such as axial load and bending, it might happen that the inelastic strains in the compressive zone, which represents the fracture process zone,  are independent of the element size, whereas on the tensile side, the width of the zones of localised strains are mesh-dependent \cite{GraJir05}.
For such a case, the adjustment of the softening modulus does not work and more advanced methods, such as nonlocal models, need to be applied.
Furthermore, mesh alignment can influence the orientation of strain profiles obtained \cite{JirGra07,GraRem07}.
In spite of these reservations, these models are used in many commercial finite element programs for modelling of fracture in concrete.

Nonlocal models, on the other hand, are based on spatial averaging of history variables in the vicinity of a material point.
Nonlocal constitutive models for concrete fracture describe the size of the average of stochastic fracture process zones independently of the size of the finite elements, as long as the finite elements are smaller than the width of the fracture process zone \cite{BazJir02, KorNguHou05, NguHou08}.
This is achieved by describing the stress at a point of the structure by means of the weighted average of state variables in the vicinity of this point.
Nonlocal models have several important advantages over the approach based on mesh-adjustment of the softening modulus.
Nonlocal approaches do not rely on the assumption that strains localise in a mesh dependent zone.
Therefore, they are capable of describing the simultaneous occurrence of distributed and localised failure mesh-independently.
Furthermore, nonlocal models have the potential to describe the fracture process, in which the fracture process zone interacts with geometric boundaries, such as corners, stress-free cracks and topological boundaries and interfaces between different materials.
Additionally, fracture patterns obtained with nonlocal models are not strongly influenced by the alignment of the finite element mesh.
However, the spatial averaging of history variables is obtained by means of an averaging operator, which is not uniquely defined for different geometric boundary conditions \cite{JirRolGra04}.
There is generally a lack of agreement, which averaging operator should be adapted and consequently the nonlocal models are currently not used in commercial finite element codes, despite their advantages over local models relying on mesh-dependent adjustment of the softening modulus.
More research is required to resolve this disagreement to utilise the full potential of nonlocal models.  

Constitutive models for fracture of concrete are often based on strain-based damage mechanics, stress-based elasto-plasticity or combined damage-plasticity frameworks.
Stress-based plasticity models are especially useful for the modelling of concrete subjected to triaxial stress states, since the yield surface corresponds at a certain stage of hardening to the strength envelope of concrete, which has a strong physical meaning \cite{Willam74, Etse94, Menetrey95, KanWil00, GraLunGyl02}.
Furthermore, the strain decomposition into elastic and plastic parts corresponds to the observed deformations in confined compression, so that unloading and path dependency can be accounted for in a realistic way.
One of the disadvantages of the stress-based plasticity models is the implicit stress evaluation procedure, where for a given strain increment the final stress state must be iterated for satisfying the yield condition.
Also, plasticity models are unable to describe the stiffness degradation observed in experiments.
On the other hand, strain based damage mechanics models are based on the concept of a gradual reduction of the elastic stiffness driven by a total strain measure.
These models are usually based on an explicit stress-evaluation algorithm, which allows for a direct determination of the stress state, without iterative calculation procedures. Furthermore, the elastic stiffness degradation observed in experiments in the case of tensile and low confined compression loading can be described. However, damage mechanics models are unable to describe irreversible deformations and are not suitable to describe the volumetric expansion \cite{Mazars84}.
Combinations of plasticity and damage are usually based on isotropic plasticity combined with either isotropic or anisotropic damage \cite{EinHouNgu07, Jef03}.
Anisotropic damage models for brittle materials, such as concrete, are often complex and, combined with plasticity, an application to structural analysis is not straightforward \cite{CarRizWil01a, HanWilCar01}.
On the other hand, combinations of isotropic damage and plasticity are widely used and many different models have been proposed in the literature.
One popular class of models relies on a combination of stress-based plasticity formulated in the effective stress space combined with a strain based damage model \cite{Ju89, LeeFen98, JasHuePijGha06, GraJir06, GraRem08, HeeJef08}.

In the present paper, both a local version based on the adjustment of softening modulus and a nonlocal version based on spatial averaging of strains of a combined damage-plasticity model is proposed, for a mesh-independent description of the fracture process of concrete.
Special focus is placed on the unloading stiffness, which for damage-plasticity is neither perfectly secant nor perfectly elastic. 
To the author's knowledge, the mesh-dependence of the unloading stiffness for damage-plasticity models was not discussed before for the case of tensorial damage-plasticity models for concrete fracture. 
However, related work was presented recently for vectorial damage-plasticity models in the context of lattice approaches for concrete fracture \cite{GraRem08}.
The present work is not meant to favorise the use of either local models based on adjustment of the softening modulus or nonlocal models. For this, a much more extensive comparison would be required, which takes into account the influence of mesh-alignment and benchmarks in which multiaxial stress-states are dominant.
Instead, the aim is to demonstrate how the two approaches can be applied to a damage-plasticity model, so that mesh-independent results are obtained. 

\section{Damage-plasticity constitutive model}
The constitutive model is based on a combination of scalar damage and stress based plasticity.
The stress-strain law is 
\begin{equation} \label{eq:general}
\boldsymbol{\sigma} = \left(1-\omega\right) \mathbf{D}_{\rm e} : \left(\boldsymbol{\varepsilon} - \boldsymbol{\varepsilon}_{\rm p}\right) = \left(1-\omega\right) \bar{\boldsymbol{\sigma}}
\end{equation}
where $\omega$ is the scalar damage parameter, ranging from 0 (undamaged) to 1 (fully damaged), $\mathbf{D}_{\rm e}$ is the elastic stiffness tensor, $\boldsymbol{\varepsilon}$ is the strain tensor, $\boldsymbol{\varepsilon}_{\rm p}$ is the plastic strain tensor and $\bar{\boldsymbol{\sigma}}$ is the effective stress tensor.

The plasticity model is based on the effective stress and thus independent of damage. 
The model is described by the yield function, the flow rule, the evolution law for the hardening variable and the loading-unloading conditions. 
The form of the yield function is
\begin{equation} \label{eq:yield}
f \left(\boldsymbol{\sigma}, \kappa \right) = F \left(\bar{\boldsymbol{\sigma}}\right) - \sigma_{\rm y} \left( \kappa \right)
\end{equation} 
where $\sigma_{\rm y}$ is the yield stress, which depends on the plastic hardening parameter $\kappa$.
The flow rule is
\begin{equation}
\boldsymbol{\varepsilon}_{\rm p} = \dot{\lambda} \dfrac{\partial g}{\partial \bar{\boldsymbol{\sigma}}}
\end{equation}
where $\dot{\lambda}$ is the rate of the plastic multiplier and $g$ is the plastic potential.
If the yield function coincides with the plastic potential ($f=g$) the flow rule is associated.

The rate of the hardening parameter $\kappa$ is
\begin{equation}
\dot{\kappa} = \left \| \dot{\boldsymbol{\varepsilon}}_{\rm p} \right \| = \dot{\lambda} \left \| \dfrac{\partial g}{\partial \bar{\boldsymbol{\sigma}}} \right \|
\end{equation}
which is the magnitude of the plastic strain rate.
The yield stress is defined as
\begin{equation}\label{eq:yieldStress}
\sigma_{\rm y} = f_{\rm t} \left( 1 + H_{\rm p} \kappa \right) 
\end{equation}
where $f_{\rm t}$ is the tensile strength. The form of $H_{\rm p}$ is discussed in detail after the description of the damage part of the model.

\begin{figure}
\begin{center}
\epsfig{file = 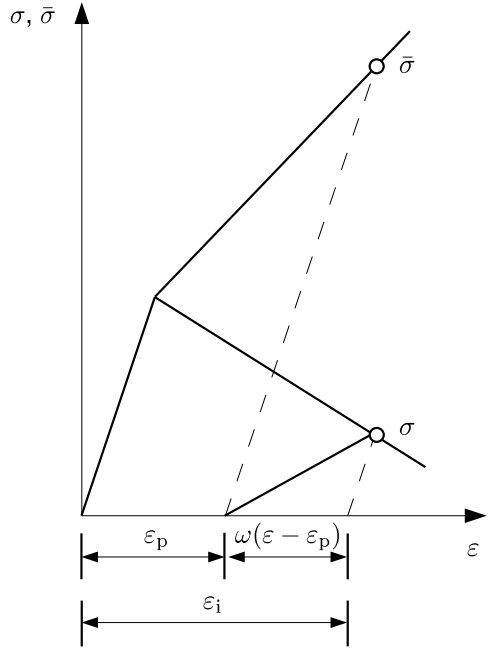, height = 8.84cm}
\end{center}
\caption{Geometrical meaning of the inelastic strain \protect $\varepsilon_{\rm i}$ for the combined damage-plasticity model. The inelastic strain is composed of a reversible \protect $\omega \left(\varepsilon - \varepsilon_{\rm p}\right)$ and irreversible part $\varepsilon_{\rm p}$.}
\label{fig:compExplain}
\end{figure}

The damage part of the present damage-plasticity model is directly linked to the plasticity model.
Only an increase of the plastic strain results in an increase of damage. However, the plasticity model is independent of damage.
The form of the damage parameter is motivated by the definition of the inelastic strain for the combined damage-plasticity model. 
The geometrical interpretation of the inelastic strain and its decomposition for uniaxial tension is shown in Fig.~\ref{fig:compExplain}.

The stress-strain law in Eq.~(\ref{eq:general}) is rewritten as 
\begin{equation}
\boldsymbol{\sigma} = \mathbf{D}_{\rm e} : \left(\boldsymbol{\varepsilon} - \left(\boldsymbol{\varepsilon}_{\rm p} + \omega \left(\boldsymbol{\varepsilon}-\boldsymbol{\varepsilon}_{\rm p}\right)\right)\right)
\end{equation}
so that the inelastic strain, which is subtracted from the total strain is
\begin{equation}\label{eq:inelastic}
\boldsymbol{\varepsilon}_{\rm i} = \boldsymbol{\varepsilon}_{\rm p} + \omega \left( \boldsymbol{\varepsilon} - \boldsymbol{\varepsilon}_{\rm p} \right)
\end{equation}
The damage history variables are chosen as the norm of the two inelastic strain components
\begin{equation}
\kappa_{\rm d1} = \left \| \boldsymbol{\varepsilon}_{\rm p} \right \|
\end{equation}
and
\begin{equation}\label{eq:kappa2}
\kappa_{\rm d2} =\max_{\tau < t} \left \| \boldsymbol{\varepsilon} - \boldsymbol{\varepsilon}_{\rm p} \right \|
\end{equation}
which are the magnitude of the plastic strain and the maximum magnitude of the elastic strain, respectively.

These history variables are related to the damage parameter for monotonic uniaxial tension, for which the uniaxial tension the yield equation (Eq.~\ref{eq:yield}) is written as
\begin{equation}
f = \sigma - \sigma_{\rm y}
\end{equation}
Thus, at the onset of yielding $\kappa_{\rm d1} = 0$ and $\kappa_{\rm d2} = \sigma_{\rm y}/E$.
A linear softening law of the form
\begin{equation} \label{eq:softening}
\sigma = f_{\rm t} \left(1 - \dfrac{\varepsilon_{\rm i}}{\varepsilon_{\rm f}}\right)
\end{equation}
was chosen, where $\varepsilon_{\rm i}$ is the inelastic strain in monotonic uniaxial tension.
According to Eqs.~(\ref{eq:inelastic})-(\ref{eq:kappa2}), this uniaxial inelastic strain is expressed by means of the two history variables $\kappa_{\rm d1}$ and $\kappa_{\rm d2}$ as
\begin{equation}\label{eq:uniInelastic}
\varepsilon_{\rm i} = \kappa_{\rm d1} + \omega \kappa_{\rm d2} 
\end{equation}
Furthermore, the stress-strain law in Eq.~(\ref{eq:general}) reduces in monotonic uniaxial tension to 
\begin{equation}\label{eq:uniaxial}
\sigma = \left( 1 - \omega \right) E \kappa_{\rm d2}
\end{equation}
Setting Eq.~(\ref{eq:softening}) equal to Eq.~(\ref{eq:uniaxial}) and using Eq.~(\ref{eq:uniInelastic}) gives
\begin{equation} \label{eq:damageParam}
 \omega = \dfrac{f_{\rm t} \kappa_{\rm d1} + \varepsilon_{\rm f} E \kappa_{\rm d2} - \varepsilon_{\rm f} f_{\rm t}}{ \kappa_{\rm d2} \left(\varepsilon_{\rm f} E - f_{\rm t} \right)}
\end{equation}
This expression is further simplified by setting
\begin{equation} \label{eq:kappa2Elastic}
\kappa_{\rm d2} = \sigma_{\rm y}/E = \dfrac{f_{\rm t}}{E} \left(1+H_{\rm p} \kappa_{\rm d1}\right)
\end{equation}
with Eq.~(\ref{eq:yield}) for $f=0$ and $\kappa_{\rm d1} = \kappa$ in Eq.~(\ref{eq:yieldStress}).
This relation exists, since the history variable $\kappa_{\rm d2}$ is defined in Eq.~(\ref{eq:kappa2}) as the maximum elastic strain, which for the case of uniaxial tension is given by Eq.~(\ref{eq:kappa2Elastic}).
Thus, the damage parameter $\omega$ is
\begin{equation}\label{eq:finalDamage}
\omega = \dfrac{f_{\rm t} E \kappa_{\rm d1} \left(1 + \varepsilon_{\rm f} H_{\rm p} \right)}{\left(1+H_{\rm p}\kappa_{\rm d1}\right) \left(\varepsilon_{\rm f} E f_{\rm t} - f_{\rm t}^2\right)}
\end{equation}
which depends only on $\kappa_{\rm d1}$, which is equal to the magnitude of the plastic strain $\kappa$.
The hardening modulus $H_{\rm p}$ influences the decomposition of the inelastic strain into a permanent and a reversible component.
The greater $H_{\rm p}$, the less is the permanent component.
A physical interpretation of the hardening modulus is given by relating it to the ratio of permanent (plastic) and total inelastic strain, which is here denoted as the unloading ratio:
\begin{equation}\label{eq:defMu}
\mu = \dfrac{\varepsilon_{\rm p}}{\varepsilon_{\rm i}}
\end{equation}
If $\mu=0$, the unloading stiffness is the secant stiffness. 
On the other hand, $\mu=1$ corresponds to the elastic unloading.
For uniaxial tension, the plastic strain can be determined analytically.
From the consistency condition 
\begin{equation}
\dot{f} = \dot{\sigma} - f_{\rm t} \dot{q} = E \left(\dot{\varepsilon} - \dot{\lambda}\right) - f_{\rm t} H_{\rm p} \dot{\lambda} = 0
\end{equation}
the plastic multiplier is
\begin{equation}\label{eq:rateLambda}
\dot{\lambda} = \dfrac{E \dot{\varepsilon}}{E + f_{\rm t} H_{\rm p}}
\end{equation}
Consequently, for monotonic loading, the total plastic strain is
\begin{equation} \label{eq:tensionEp}
\varepsilon_{\rm p} = \dfrac{E \varepsilon - f_{\rm t}}{E+f_{\rm t} H_{\rm p}}
\end{equation}
Hence, the ratio of the plastic strain in Eq.~(\ref{eq:tensionEp}) and the total inelastic strain is
\begin{equation}\label{eq:initMu}
\mu = \dfrac{\varepsilon_{\rm p}}{\varepsilon_{\rm i}} = \dfrac{E \varepsilon - f_{\rm t}}{\varepsilon_{\rm i} \left(E + f_{\rm t} H_{\rm p}\right)}
\end{equation}
For the case that the material approaches the fully damaged stage ($\omega \rightarrow 1$), the total strain is equal to the inelastic strain, since there are no elastic strains remaining, resulting in $\varepsilon_{\rm i} = \varepsilon = \varepsilon_{\rm f}$.
Thus, Eq.~(\ref{eq:initMu}) is written as
\begin{equation}
\mu = \dfrac{E \varepsilon_{\rm f} - f_{\rm t}}{\varepsilon_{\rm f} \left(E + f_{\rm t} H_{\rm p}\right)}
\end{equation}
from which the hardening modulus $H_{\rm p}$ is determined as
\begin{equation} \label{eq:hardModulus}
H_{\rm p} = \dfrac{E \varepsilon_{\rm f} \left(1 - \mu \right) - f_{\rm t}} {\mu \varepsilon_{\rm f} f_{\rm t}}
\end{equation}For $\mu =1$, which corresponds to the case of elastic unloading, the hardening modulus is $H_{\rm p} = -1/\varepsilon_{\rm f}$.
The other limit, $\mu \rightarrow 0$, corresponds to a hardening modulus $H_{\rm p} \rightarrow \infty$.
Furthermore, $H_{\rm p} = 0$ corresponds to $\mu = 1-\varepsilon_0/\varepsilon_{\rm f}$.
The present interpretation of the hardening modulus $H_{\rm p}$ is based on uniaxial considerations and the assumption that the material is close to a fully damaged stage. 
For the linear softening, this stage is reached when $\varepsilon = \varepsilon_{\rm f}$. 
For an exponential softening law of the form 
\begin{equation}
\sigma = f_{\rm t} \exp \left(-\dfrac{\varepsilon_{\rm i}}{\varepsilon_{\rm f}}\right)
\end{equation}
which was used for the analyses of the three-point bending test, the material is not fully damaged at $\varepsilon = \varepsilon_{\rm f}$. 
Therefore, for exponential softening the parameter $\mu$ does not represent the ratio of permanent and total inelastic strain.
However, the unloading ratio is independent of the element size used.
For multiaxial stress-states the above derivations need to be extended and generalised. 
Some progress was recently made in \cite{GraRem08}, where an approach similar to the one presented here was applied to a vectorial constitutive model for tension, shear and compression.

The plastic-damage model possesses two dissipation mechanisms.
Let us define the free energy as
\begin{equation}
\rho \psi \left(\sbf{\veps}, \sbf{\veps}_{\rm p}, \omega \right) = \dfrac{1}{2} \left( 1- \omega\right) \left(\sbf{\veps} - \sbf{\veps}_{\rm p}\right) : \mbf{D}_{\rm e} : \left(\sbf{\veps} - \sbf{\veps}_{\rm p}\right)
\end{equation}
Here, $\rho$ is the density (specific mass), $\psi$ is the Helmholtz free energy per unit mass.
Furthermore,  isothermal processes are considered, so that the temperature remains constant and is not explicitly listed among the state variables.
The dissipation rate per unit volume is defined as
\begin{equation}
\dot{D} = \sbf{\sigma}: \dot{\sbf{\veps}} - \rho \dot{\psi} = \sbf{\sigma}:\dot{\sbf{\veps}} - \left(\pard{\rho\psi}{\sbf{\veps}}{\dot{\sbf{\veps}}} + \pard{\rho \psi}{\sbf{\veps}_{\rm p}} \dot{\sbf{\veps}}_{\rm p} + \pard{\rho \psi}{\omega} \dot{\omega}\right)
\end{equation}

\begin{figure}
\begin{center}
\begin{tabular}{cc}
\epsfig{file = 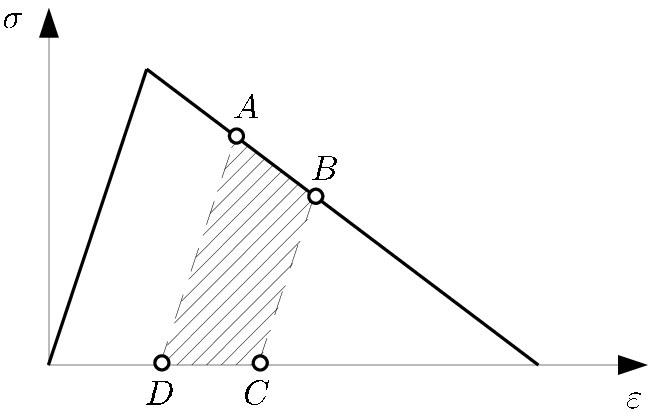, height = 4.1cm} & \epsfig{file = 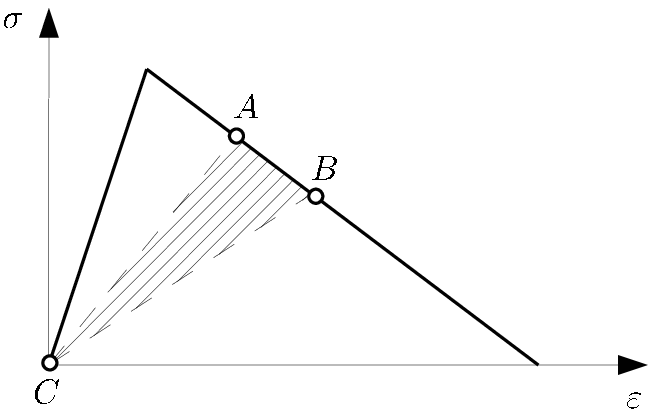, height = 4.1cm} \\
(a) & (b)
\end{tabular}
\end{center}
\caption{Dissipation for (a) pure plasticity and (b) pure damage. The hatched areas denote the dissipation for softening from $A$ to $B$.} 
\label{fig:dissExplain}
\end{figure}

The derivative of the free energy with respect to the strain results in the stress-strain relation
\begin{equation}
\pard{\rho \psi}{\sbf{\veps}} \dot{\sbf{\veps}} = \left(1-\omega\right) \left(\sbf{\veps} - \sbf{\veps}_{\rm p}\right): \mbf{D}_{\rm e} : \dot{\sbf{\veps}} = \sbf{\sigma} : \dot{\sbf{\veps}}
\end{equation}
Therefore, the dissipation rate reduces to
\begin{equation} \label{eq:dissipation}
\dot{D} =  \dot{D}^{\rm p} + \dot{D}^{\rm e} = \pard{\rho \psi}{\sbf{\veps}_{\rm p}} \dot{\sbf{\veps}}_{\rm p} + \pard{\rho \psi}{\omega} \dot{\omega}
\end{equation}
where the term
\begin{equation}
\dot{D}^{\rm p} = \pard{\rho \psi}{\sbf{\veps}_{\rm p}} \dot{\sbf{\veps}}_{\rm p} = \left(1-\omega\right) \left(\sbf{\veps} - \sbf{\veps}_{\rm p}\right) : \mbf{D}_{\rm e}: \dot{\sbf{\veps}}_{\rm p}
\end{equation}
and
\begin{equation}
\dot{D}^{\rm d} = \pard{\rho \psi}{\omega} \dot{\omega} = \dfrac{1}{2} \left(\sbf{\veps} - \sbf{\veps}_{\rm p}\right) : \mbf{D}_{\rm e} : \left(\sbf{\veps} - \sbf{\veps}_{\rm p}\right) \dot{\omega}
\end{equation}
is the plastic and and damage dissipation, respectively.
The two different dissipation mechanisms are illustrated in Fig.~\ref{fig:dissExplain}a~and~b for pure damage and pure plasticity, respectively.

\subsection{Mesh-adjusted softening modulus}
The first approach to obtain mesh-independent results is based on the idea to adjust the softening modulus with respect to the element size.
This approach, which is often called crack band approach \cite{BazOh83}, is very common for pure damage mechanics or pure plasticity models, where the aim is to obtain a mesh-independent softening response. 
For the combined damage-plasticity model, it is also important to ensure a mesh-independent unloading stiffness.
Therefore, both the plasticity and damage part of the model are made dependent of the element size.
This is achieved by replacing in Eq.~(\ref{eq:damageParam}) and Eq.~(\ref{eq:hardModulus}) $\varepsilon_{\rm f}$ with $w_{\rm f}/h$, where $w_{\rm f}$ is the inelastic displacement at which the stress becomes zero and $h$ is the size of the finite element element.
Consequently, the damage parameter for the crack band approach is
\begin{equation}
\omega = \dfrac{ f_{\rm t} E \kappa_{\rm d1} \left(1+\dfrac{w_{\rm f}}{h} H_{\rm p}\right)}{\left(1+H_{\rm p} \kappa_{\rm d1}\right) \left(\dfrac{w_{\rm f}}{h} E f_{\rm t} - f_{\rm t}^2\right)}
\end{equation}
and the hardening modulus is
\begin{equation}
H_{\rm p} = \dfrac{E \dfrac{w_{\rm f}}{h} \left(1-\mu \right) - f_{\rm t}}{\mu \dfrac{w_{\rm f}}{h} f_{\rm t}}
\end{equation}
For $\mu = 1$ a pure plastic model is obtained for which the hardening modulus is $H_{\rm p} = -h/w_{\rm f}$.
The smaller the element size $h$, the bigger is the hardening modulus.

\subsection{Nonlocal model}
Integral-type nonlocal models overcome the problems of the local softening models by spatial averaging of suitable state variables.
For plasticity combined with nonlocal damage mechanics, the damage history variable $\kappa_{\rm d1}$ in Eq.~(\ref{eq:finalDamage}) is replaced by
\begin{equation} \label{eq:average}
\bar{\kappa}_{\rm d1} \left(\mbf{x}\right) = \int_V  \alpha\left(\mbf{x}, \mbf{s}\right) \kappa_{\rm d1} \left(\mbf{s}\right)  \mbox{d} \mbf{s}
\end{equation}
so that the damage parameter $\omega$ is then determined by means of $\bar{\kappa}_{\rm d1}$ instead of $\kappa_{\rm d1}$.
All other equations of the plastic damage model remain the same, i.e. no adjustment of the softening modulus with respect to the element size is performed. 
The plasticity part of the constitutive model remains local. The nonlocal approach is only applied to the damage part.
In Eq.~(\ref{eq:average}), $\alpha \left(\mbf{x}, \mbf{s}\right)$ is a nonlocal weight function that describes the interaction between points $\mbf{x}$ and $\mbf{s}$, which decays with increasing distance between these points. 
Here, $\mathbf{x}$ is the point at which the nonlocal history variable is evaluated and $\mathbf{s}$ is a point in its vicinity.
The weight function is usually non-negative and normalised such that
\begin{equation}
\int_{V} \alpha\left(\mbf{x}, \mbf{s} \right)\mbox{d}\mbf{s} = 1 \hspace{1cm} \mbox{for all $\mbf{x}$ in V}
\end{equation}
This can be achieved by setting
\begin{equation}
\alpha \left(\mbf{x},\mbf{s}\right) = \dfrac{\alpha_{\infty} \left(||\mbf{x} - \mbf{s}||\right)}{\int_V \alpha_{\infty} \left(||\mbf{x} - \sbf{\zeta}||\right) \mbox{d} \sbf{\zeta}}
\end{equation}
Typical weighting functions $\alpha_{\infty}$ are the Gauss-like function or the bell-shaped truncated polynomial function.
The bell-shaped function
\begin{equation}
\alpha_{\infty}^{\mbox{bell}}(r) = \left\{\begin{array}{ll}
\dfrac{1}{c} \left(1 - \dfrac{r^2}{R^2}\right)^2  & \mbox{if $r \leq R$}\\
0 & \mbox{if $r \geq R$}
\end{array} \right.
\end{equation}
which has a bounded support, i.e. it vanishes at distances $r = \|\mathbf{x}-\mathbf{s}\|$ exceeding the interaction radius $R$.
The scaling factor is $c=16R/15$ and $c = \pi R^2/3$ for the one and two-dimensional case, respectively \cite{Rol03}.
The present combination of local plasticity with nonlocal damage is computationally more efficient than standard integral-type nonlocal plasticity models, since the plasticity part, which is implicit, remains local and only the equivalent strain in the explicit damage part is averaged \cite{Rol03, JirGra04,GraJir05}.
In \cite{JirGra04,GraJir05} it was concluded that mesh-independent results can only be obtained, if the hardening modulus $H_{\rm p}$ of the plasticity model is greater than the critical hardening modulus of the material. 
Therefore, only $\mu$-values corresponding to $H_{\rm p} \geq 0$ are considered in the present study.

\section{Constitutive response}
\begin{figure}
\begin{center}
\begin{tabular}{cc}
\epsfig{file = 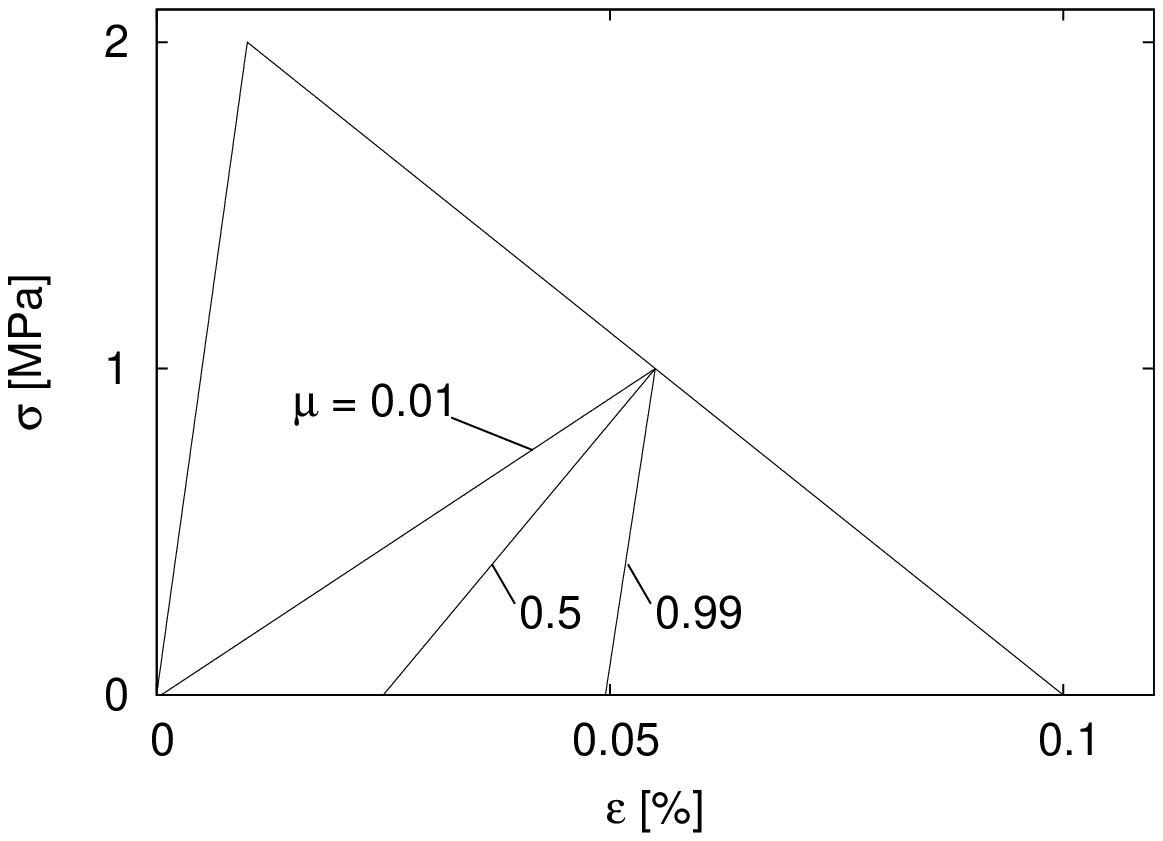, width=8.cm} & \epsfig{file = 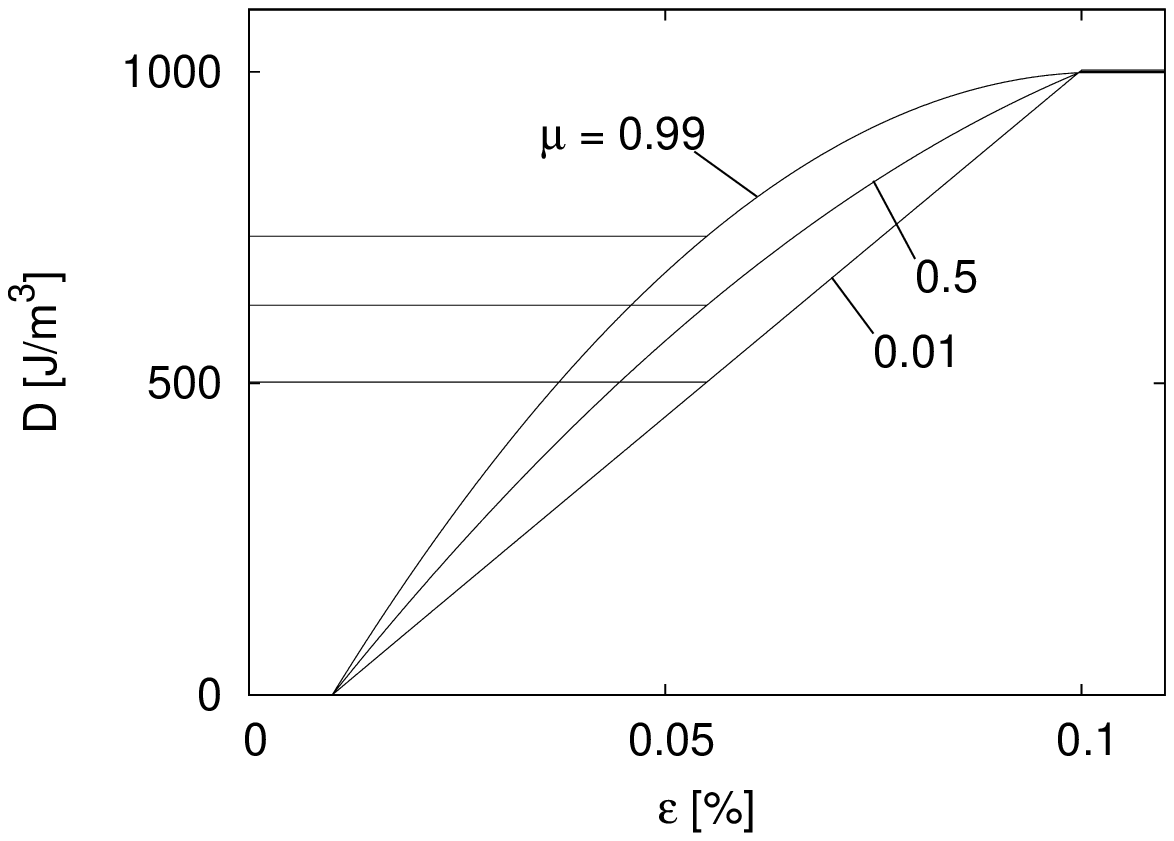, width=8.cm}\\
(a) & (b)
\end{tabular}
\caption{Constitutive response: (a)~Stress-strain curves for $\mu=0.01$, $\mu=0.5$ and $\mu=0.99$, (b)~ Dissipation versus strain for $\mu=0.01$, $\mu=0.5$ and $\mu=0.99$.}
\label{fig:hpConst}
\end{center}
\end{figure}

For illustration of the above concepts, the interaction of damage and plasticity is here studied for the case of softening in uniaxial tension.
The material properties are chosen as $E= 20$~GPa, $f_{\rm t} = 2$~MPa and $\veps_{\rm f}=0.001$. 
In Fig.~\ref{fig:hpConst}a, the stress-strain relations are shown for $\mu=0.99,0.5$~and~$0.01$.
For $\mu=0.99$ ($H_{\rm p} \approx -\dfrac{1}{\veps_{\rm f}}$), the damage variable $\omega$ in Eq.~(\ref{eq:finalDamage}) is equal to zero and a plastic response ($\veps_{\rm i} \approx \veps_{\rm p}$) is obtained.
On the other hand, for $\mu \rightarrow 0$ ($H_{\rm p} \rightarrow \infty$) the plastic multiplier in Eq.~(\ref{eq:rateLambda}) is zero and a damage response with $\veps_{\rm i} \approx \omega \veps$ is obtained.
The evolution of the dissipation determined from Eq.~(\ref{eq:dissipation}) for the three values of $\mu$ is shown in Figure~\ref{fig:hpConst}b.
For the linear softening law in Eq.~(\ref{eq:yieldStress}), the dissipation evolution of the plasticity model is nonlinear, whereas for the damage model a linear dissipation evolution is obtained.
The dissipated energy per unit volume for the present linear softening law is determined to $g_{\rm F} = \dfrac{1}{2} f_{\rm t} \veps_{\rm f}$.

\section{One-dimensional damage-plasticity response}

\begin{figure}
\begin{center}
\epsfig{file=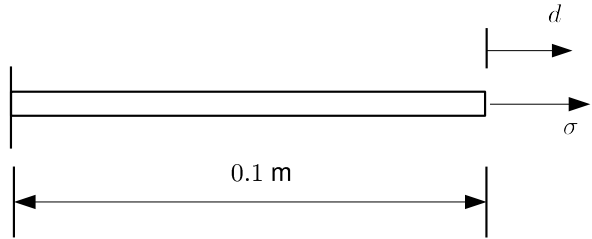, width = 8.cm}
\caption{Geometry of the uniaxial tensile specimen.}
\label{fig:figBar}
\end{center}
\end{figure}

\begin{figure}
\begin{center}
\begin{tabular}{cc}
\epsfig{file=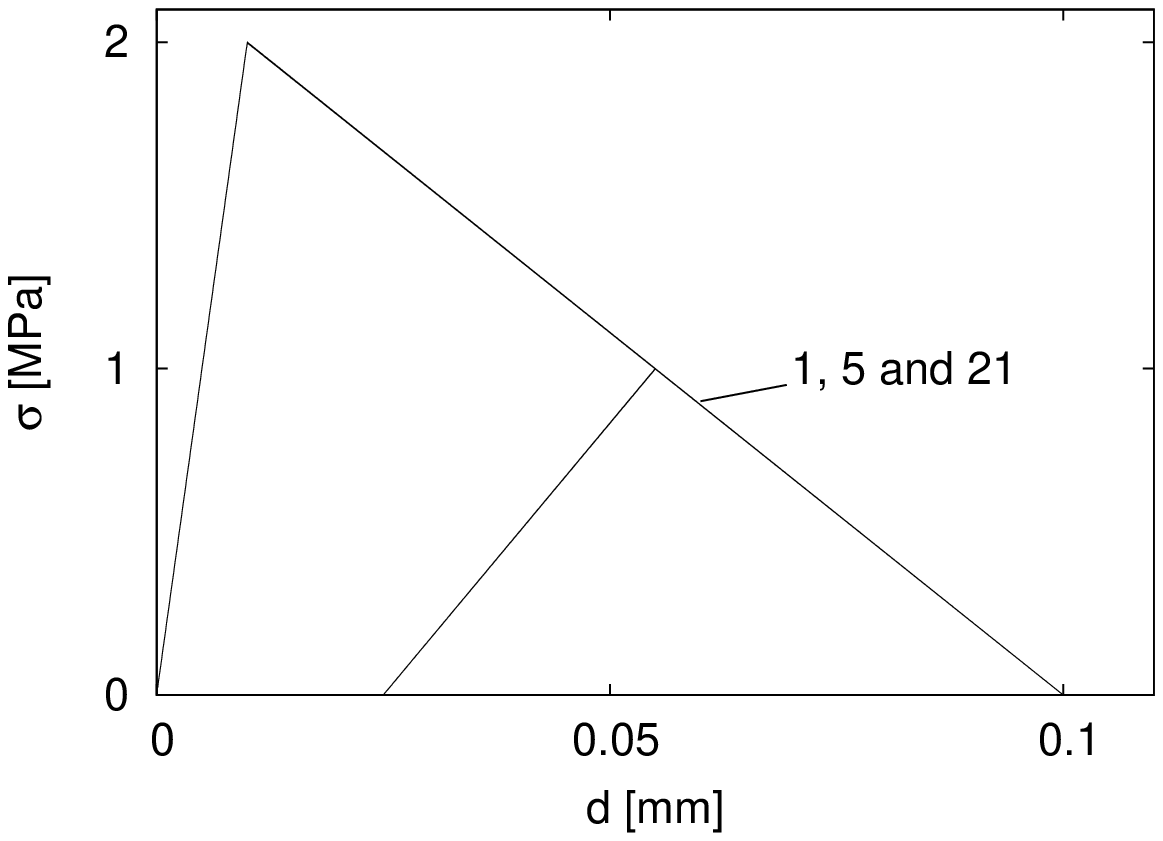, width = 8.cm} & \epsfig{file=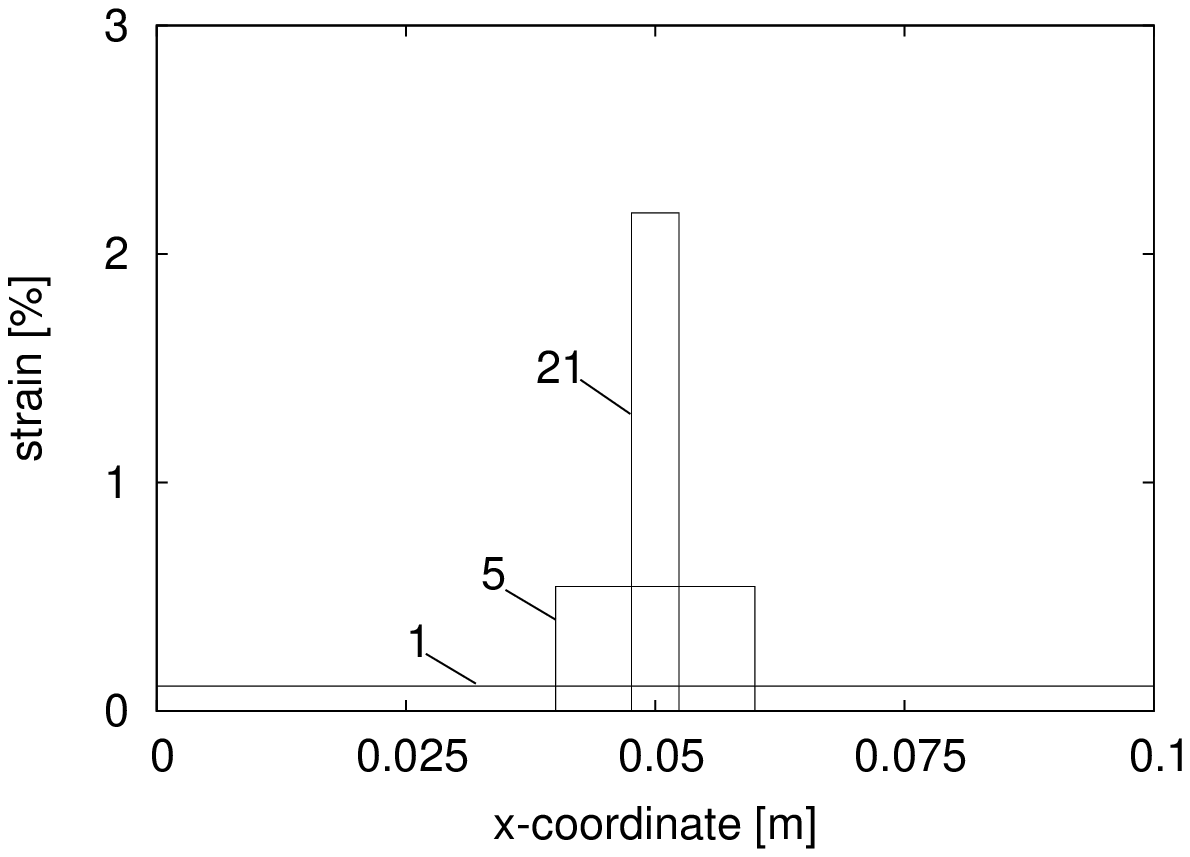, width = 8.cm}
\end{tabular}
\caption{Effect of mesh refinement on the numerical results: a) load-displacement diagrams, b) strain profiles.}
\label{fig:crackBandLd}
\end{center}
\end{figure}

Two ways of resolving the issue of mesh dependency associated with local softening constitutive models, namely mesh-adjusted softening modulus and nonlocal  damage models, are discussed further.
The two approaches are applied to a bar subjected to uniaxial tension shown in Figure~\ref{fig:figBar}. 
The material parameters are the same as used in the previous sections. 
The load-displacement curves for three mesh-sizes with 1, 5 and 21 elements are shown in Figure~\ref{fig:crackBandLd}a. 
The element located in the middle of the bar of the meshes is weakened by $0.1 \%$.
The strain profiles are shown in Figure~\ref{fig:crackBandLd}b.
The load-displacement curves are independent of the mesh-size. 
However, the strain profiles depend on the mesh size.

\begin{figure}
\begin{center}
\begin{tabular}{ll}
\epsfig{file=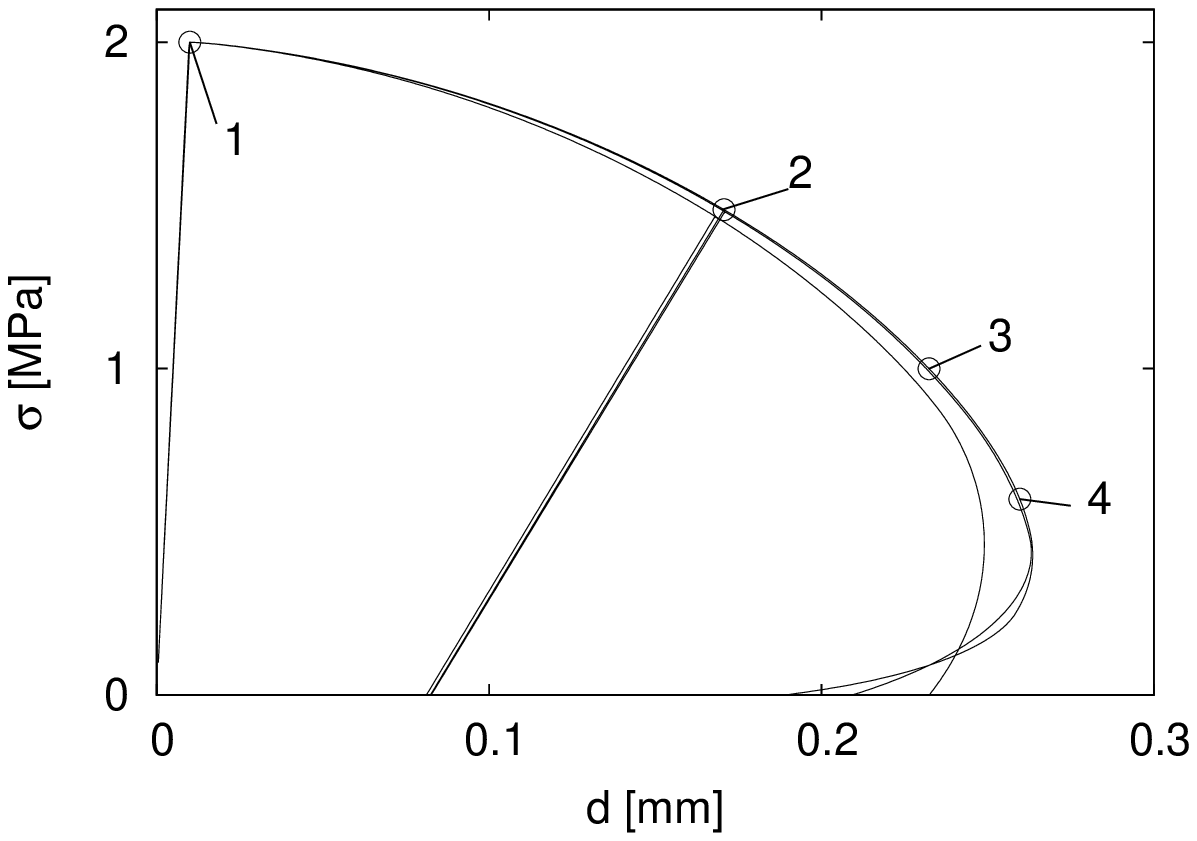, width=8cm} & \epsfig{file=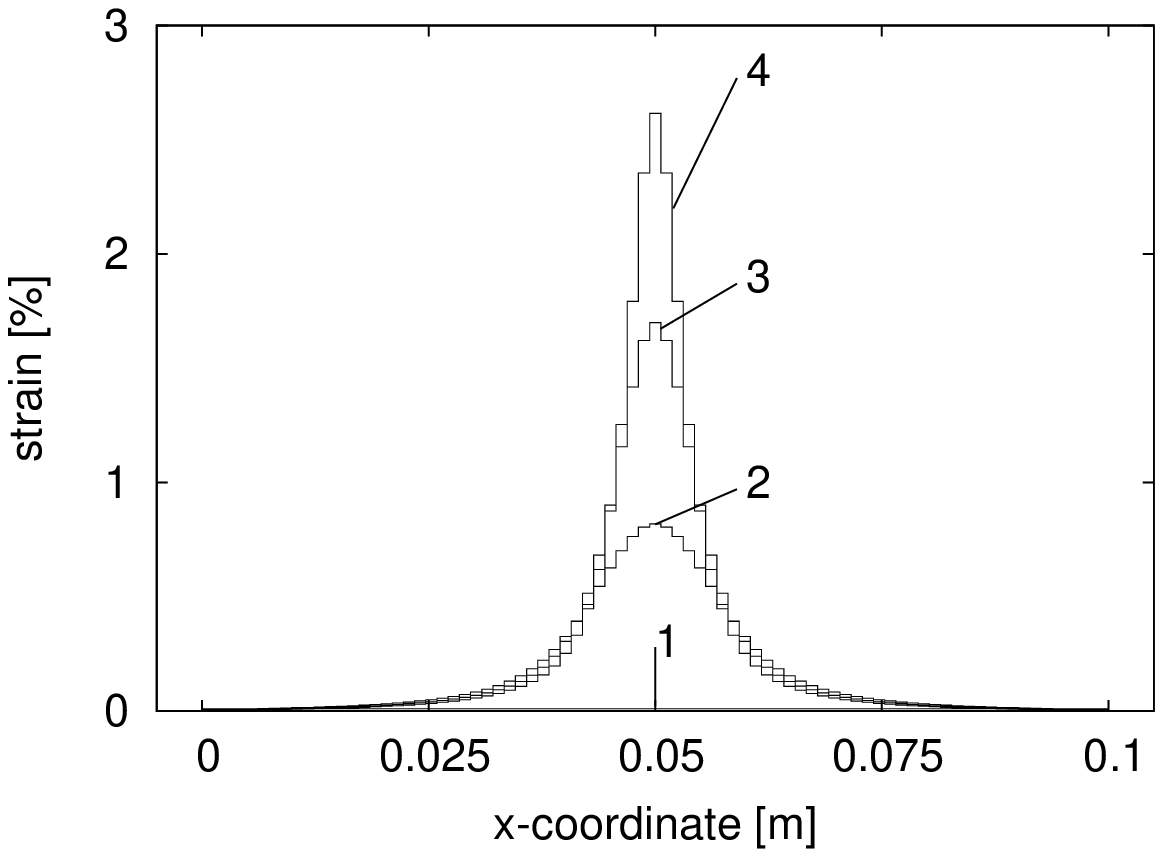,width=8cm}
\end{tabular}
\caption{Nonlocal plastic-damage model: (a) Load-displacement curves for 21, 41 and 81 elements and b) evolution of the strain distribution for 81 elements at four stages of analysis.}
\label{fig:nonlocalLd}
\end{center}
\end{figure}

The nonlocal model is applied to the same uniaxial problem which was analysed in the previous section with the mesh-adjusted softening modulus.
Three different meshes with 21, 41 and 81 elements were used.
A section of 0.005~m in the middle of the bar was weakened (Figure~\ref{fig:nonlocalLd}) by $0.1 \%$.
The nonlocal radius was chosen as $R=0.01$, which is approximately twice the element size of the coarsest mesh.
The linear softening curve from Equation~(\ref{eq:yieldStress}) with $\veps_{\rm f} = 0.002$ was used.
The other material parameters were the same as in the section presenting the constitutive response.
The unloading ratio was chosen to $\mu=0.5$.

\begin{figure}
\begin{center}
\epsfig{file=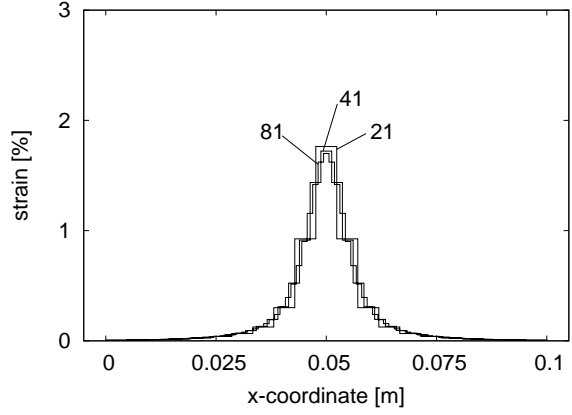, width=8cm} 
\caption{Nonlocal plastic-damage model: a) Strain distribution for 21,41 and 81 elements at stage 3 marked in Fig.~\protect\ref{fig:nonlocalLd}a.}
\label{fig:nonlocalStrain}
\end{center}
\end{figure}

\begin{figure}
\begin{center}
\begin{tabular}{ll}
\epsfig{file=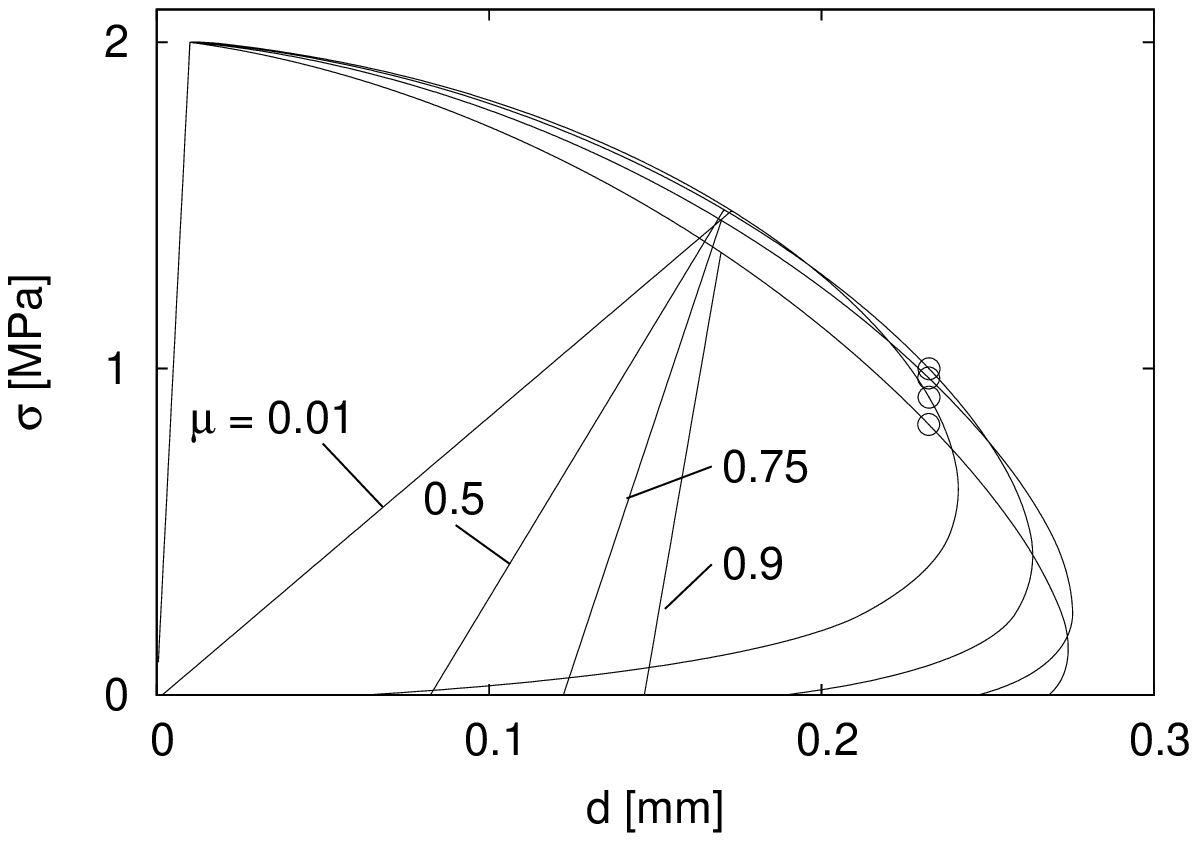,width=8cm} & \epsfig{file=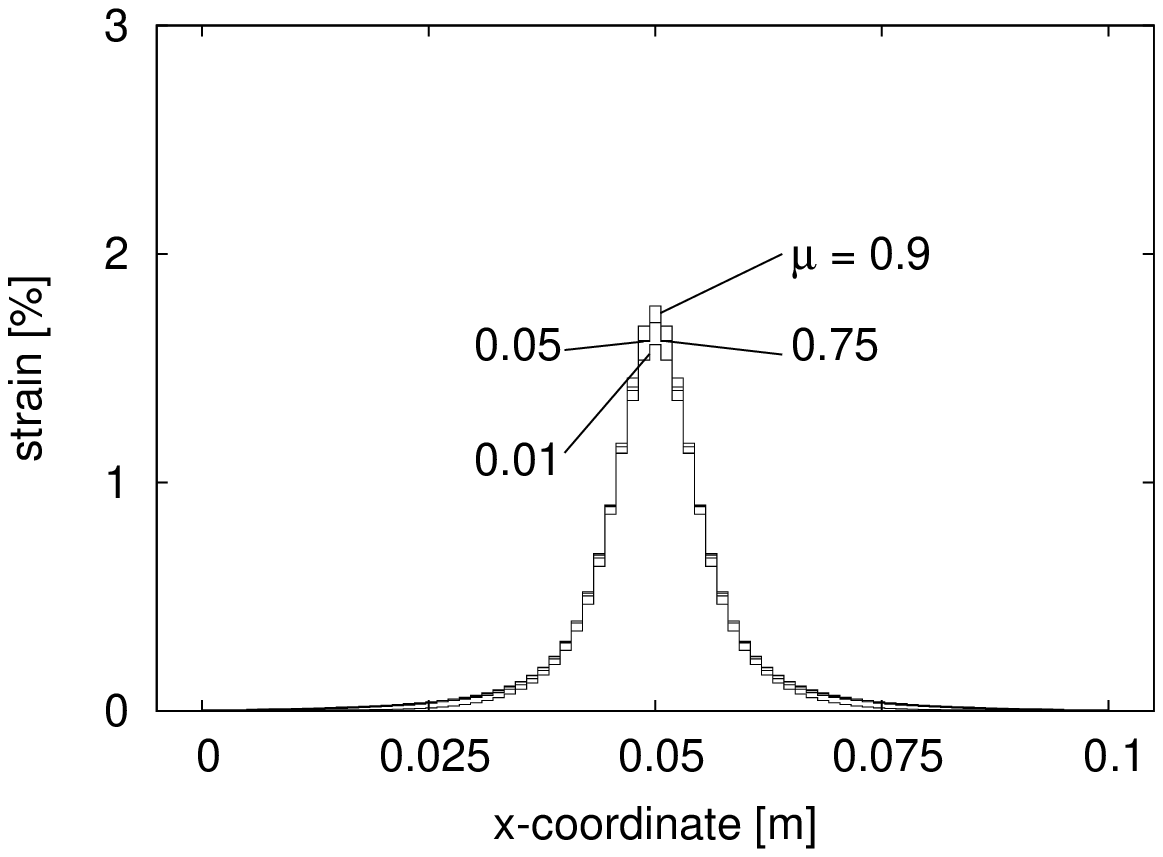, width=8cm}
\end{tabular}
\caption{Nonlocal plastic-damage model: a) Load-displacement curve for ratios $\mu=0.01, 0.5, 0.75, 0.9$ and b) strain distribution for the same four ratios at the stage marked in the load-displacement curves.}
\label{fig:nonlocalRatio}
\end{center}
\end{figure}

The load-displacement curves are shown in Figure~\ref{fig:nonlocalLd}a and demonstrate an almost mesh-independent result.
Furthermore, the strain profiles in Figure~\ref{fig:nonlocalStrain} are almost identical for the three meshes.
However, the load-displacement curves are nonlinear, although a linear local softening curve was used in Eq.~(\ref{eq:softening}).
The nonlinear load-displacement response is explained by the shrinking of the zone of localised strains (Fig.~\ref{fig:nonlocalLd}b). 
With decreasing load, the strain localises further, so that the load-displacement curve becomes steeper.
Furthermore, the influence of the ratio on the fracture process zone and load-displacement curves is investigated. 
For the nonlocal model, the damage part is nonlocal and the plasticity is local. To ensure mesh-independent results, the hardening modulus $H_{\rm p}$ must be greater than zero. 
This corresponds for the chosen model parameters to the condition $\mu \geq 0.9$.
Four unloading ratios of $\mu=0.01, 0.5, 0.75, 0.9$ were used for the analysis with the fine mesh. The load-displacement curves and strain profiles are shown in Fig.~\ref{fig:nonlocalRatio}.
The softening branch of the load-displacement curves differs slightly. However, the strain profiles in Fig.~\ref{fig:nonlocalRatio}b are almost independent of the parameter $\mu$.

\section{Three point-bending test}
\begin{figure}
\begin{center}
\epsfig{file=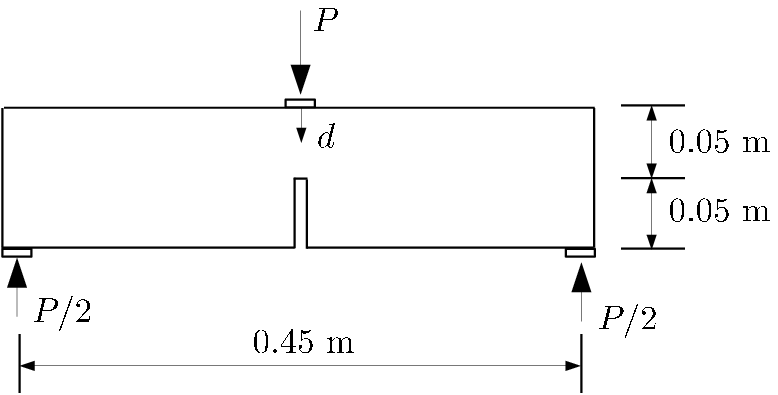, width=8cm}\vspace{1cm}\\
\epsfig{file=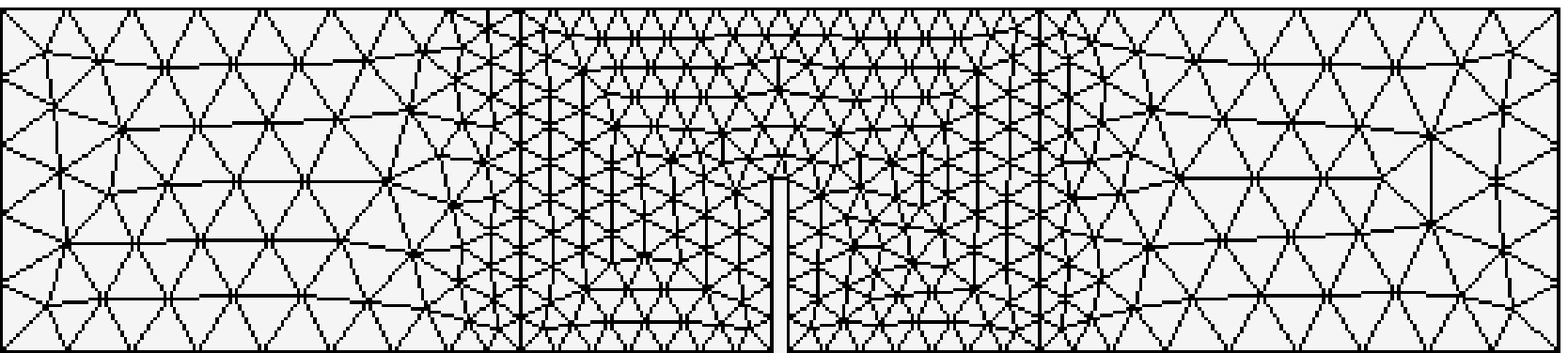, width=8cm} 
\caption{Geometry and setup of the three-point bending test.}
\label{fig:3PBTGeometry}
\end{center}
\end{figure}

\begin{figure}
\begin{center}
\epsfig{file=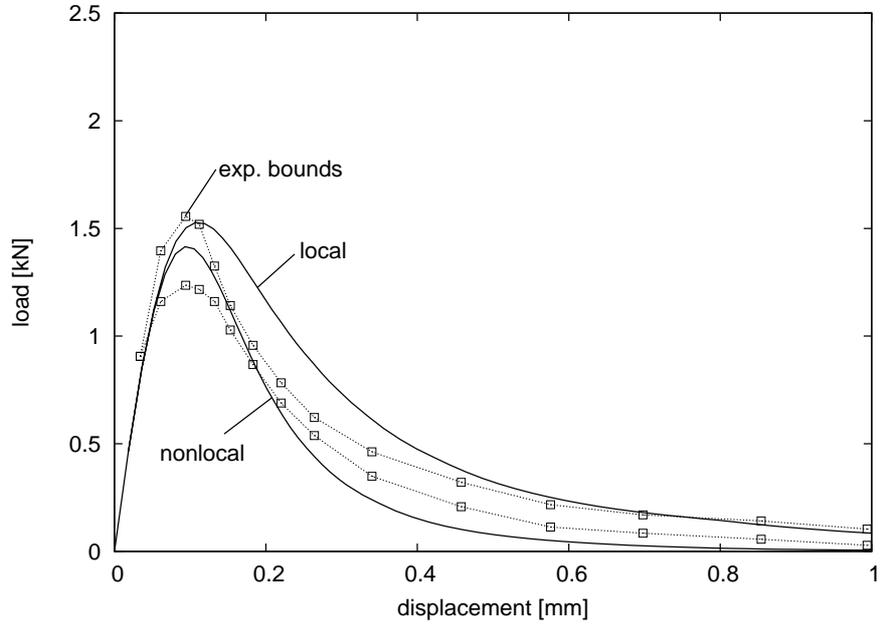,width=12cm}
\caption{Comparison of the load-displacement cuvrves obtained from the two modelling approaches and the experimental results reported in \protect \cite{KorRei83}.}
\label{fig:3PBTExpComp}
\end{center}
\end{figure}

\begin{figure}
\begin{center}
\epsfig{file=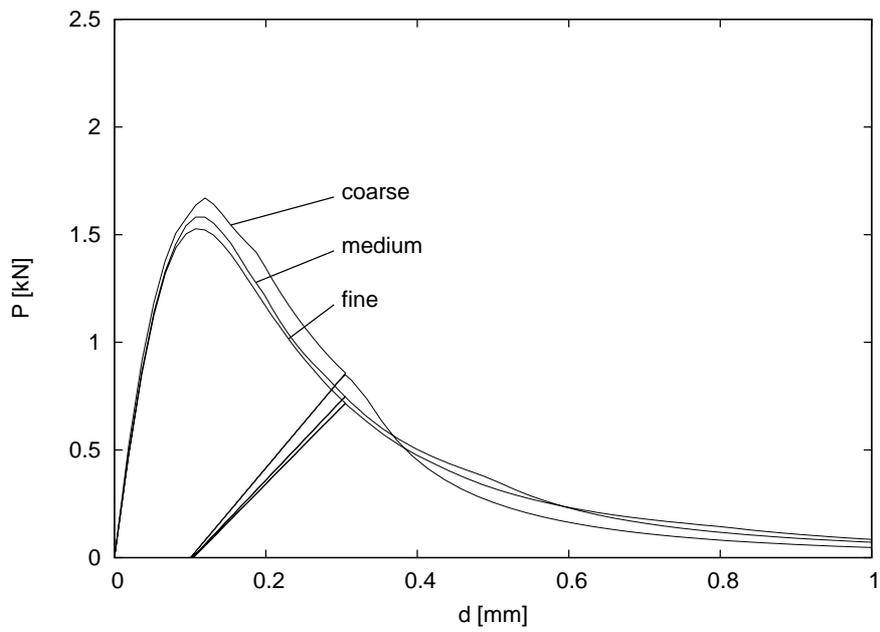, width=12cm}
\caption{Three point bending test: Load-displacement curves obtained with the mesh-adjusted softening modulus for three mesh sizes with $\mu=0.5$.}
\label{fig:fig3PBTSmearedLdMesh}
\end{center}
\end{figure}

The softening modulus adjustment and nonlocal approach are here applied to the analysis of the three point bending test, for which the experimental results were reported in \cite{KorRei83}.
For this analysis, the two-dimensional Rankine yield surface with an associated flow rule was used for the plasticity part of the damage-plasticity model \cite{Rol03}. The hardening law of the plasticity part was chosen according to Eq.~(\ref{eq:hardModulus}).
For the damage part, an exponential softening law was chosen.
The geometry and the setup of the three point bending test is shown in Fig.~\ref{fig:3PBTGeometry}.
Three finite element meshes with element size $7.5, 5$~and~$2.5$~mm in the central part of the specimen were used. 
The coarse mesh is shown in Fig.~\ref{fig:3PBTGeometry}.
The three point bending test was analysed with the damage-plasticity model using the mesh-adjusted softening modulus and the nonlocal model.
The model parameters  for the mesh-adjusted softening modulus were chosen as $E=20$~GPa, $f_{\rm t}=2.4$~MPa, $w_{\rm f} = 0.000047$~m.
Furthermore, the Poisson's ratio required for the two-dimensional analysis was chosen as $\nu = 0.2$.
For the nonlocal approach the critical crack opening $w_{\rm f}$ was replaced by the critical strain $\varepsilon_{\rm f} = 0.001025$ and the nonlocal radius is set to $R=0.025$~m.
These parameters resulted in satisfactory results for a damage-plasticity approach reported in \cite{GraJir05}, in which damage was determined by the plastic strain and the plasticity model was chosen to be perfect plastic ($H_{\rm p} = 0$), which corresponds to $\mu = 1-f_{\rm t}/\left(E \varepsilon_{\rm f} \right) = 0.883$.  
In Fig.~\ref{fig:3PBTExpComp} the present two modelling approaches with $\mu=0.883$ are compared to the experimental results reported in \cite{KorRei83}.
The results obtained with the two modelling approaches are in satisfactory agreement with the experimental results.
In the following sections the mesh-dependence and the influence of the unloading ratio $\mu$ are investigated.
For the mesh-adjusted softening modulus, the load-displacement curves for the three meshes are shown in Fig.~\ref{fig:fig3PBTSmearedLdMesh}a.
The results are almost independent of the size of the elements. 
\begin{figure}
\begin{center}
\epsfig{file=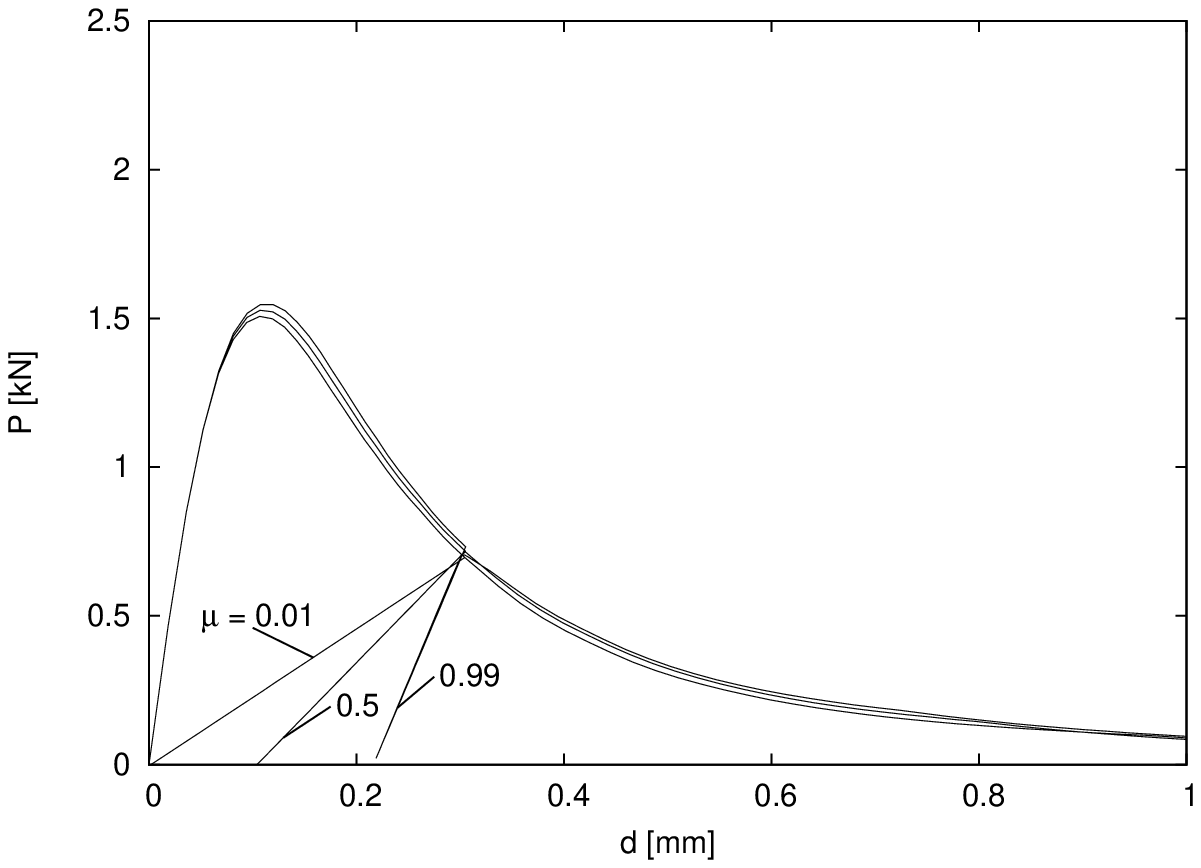, width=12cm}
\caption{Three point bending test: Load-displacement curves obtained with the mesh-adjusted softening modulus for three ratios of $\mu=0.01, 0.5$ and $0.99$ for the fine mesh.}
\label{fig:fig3PBTSmearedLdRatio}
\end{center}
\end{figure}
Furthermore, the load-displacement curves for the ratios $\mu=0.01, 0.5$~and~$0.9$ are presented for the fine mesh in Fig.~\ref{fig:fig3PBTSmearedLdRatio}b.
The unloading part of the load-displacement curves differs depending on the chosen unloading ratio $\mu$.
However, the softening part is almost independent of this ratio.
For the nonlocal model, the load-displacement curves for the three meshes are shown in Fig.~\ref{fig:fig3PBTNonlocalLdMesh} for $\mu=0.5$.
Again, the load-displacement curves are independent of the element size.
However, the maximum load is considerably higher than for $\mu=0.883$, for which the nonlocal model was calibrated to match the experimental results. 
This overestimation of the peak load is further investigated by varying the unloading ratio $\mu$.
The load-displacement curves for $\mu = 0.5, 0.8$ and $0.883$ are shown in Fig.~\ref{fig:fig3PBTNonlocalLdRatio}.
\begin{figure}
\begin{center}
\epsfig{file=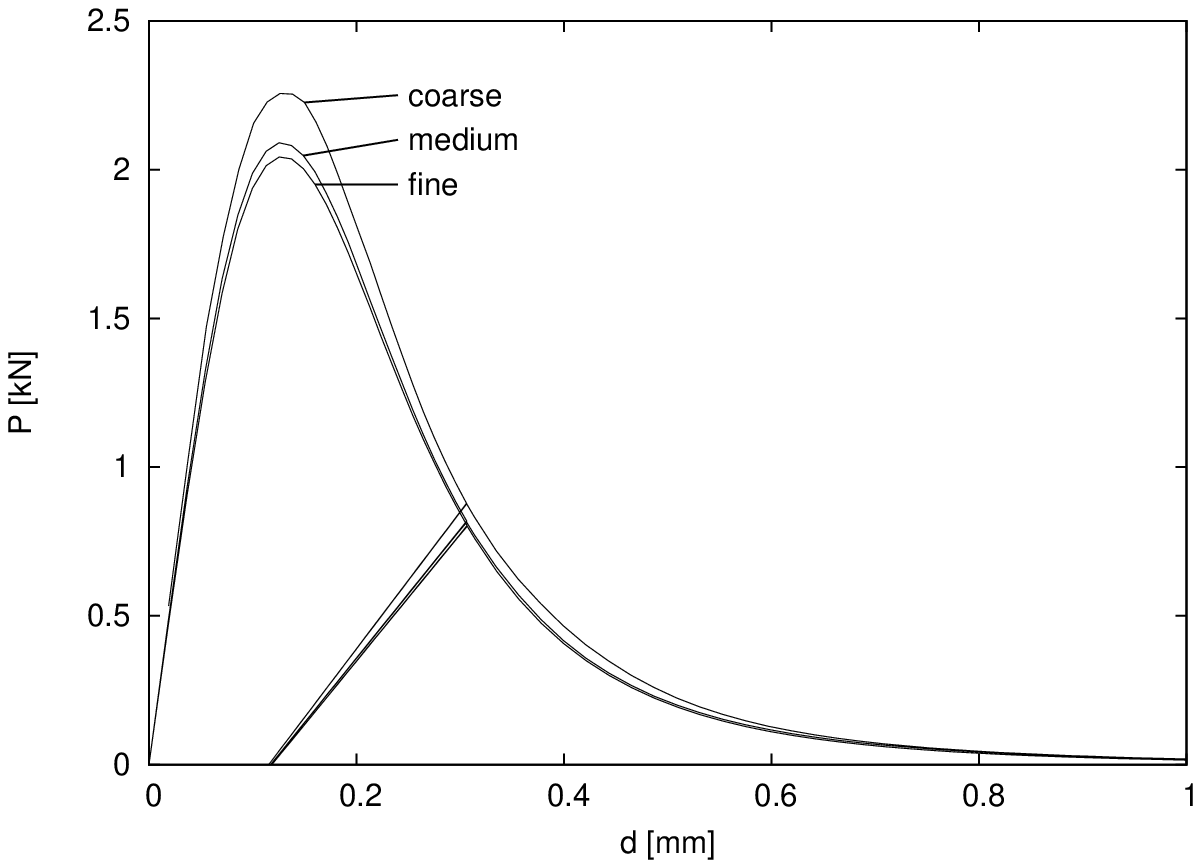, width=12cm}
\caption{Three point bending test: Load-displacement curves obtained with the nonlocal model for three mesh sizes with $\mu=0.5$.}
\label{fig:fig3PBTNonlocalLdMesh}
\end{center}
\end{figure}
\begin{figure}
\begin{center}
\epsfig{file=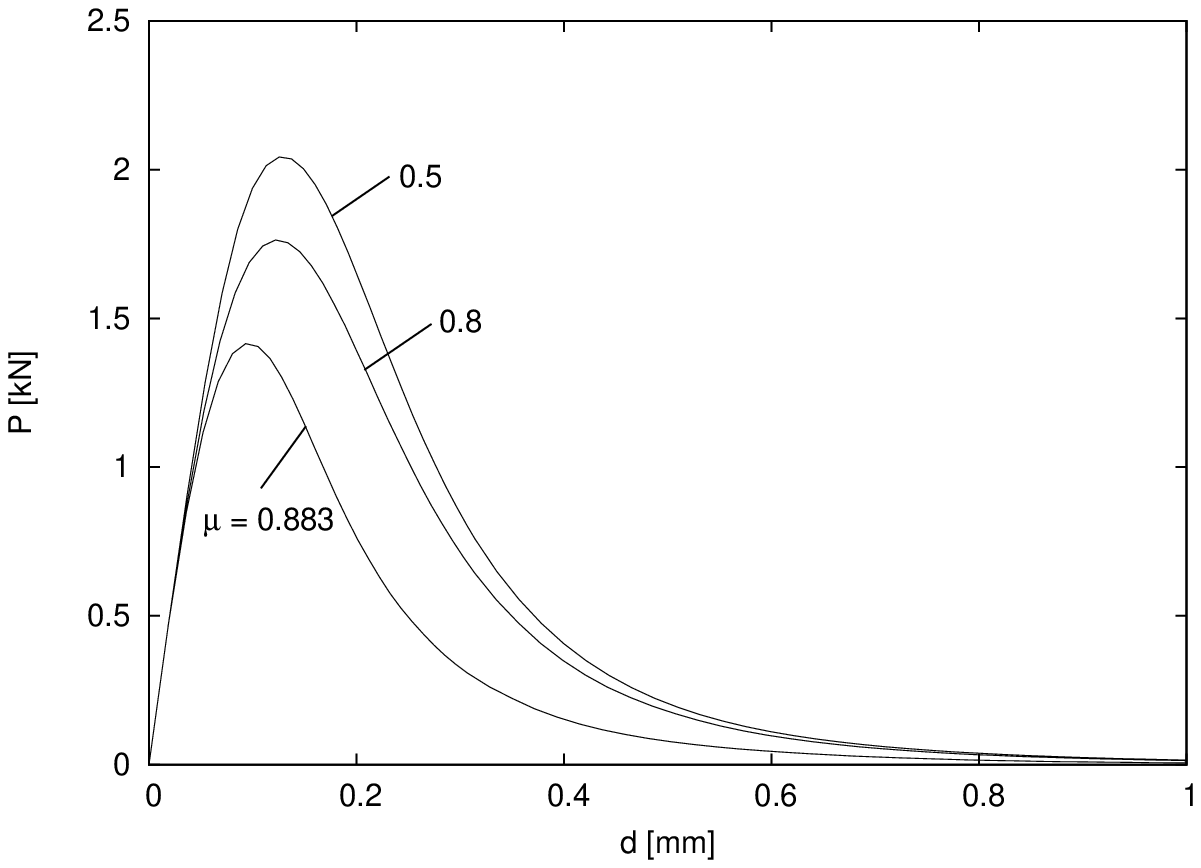, width=12cm}
\caption{Three point bending test: Load-displacement curves obtained with the nonlocal model for three ratios of $\mu=0.5, 0.8$ and $0.883$ for the fine mesh.}
\label{fig:fig3PBTNonlocalLdRatio}
\end{center}
\end{figure}
The smaller the $\mu$, the greater is the peak load.
This strong dependence on the unloading ratio $\mu$ is caused by the presence of the notch in the problem considered.
The nonlocal history variable of the damage model, which is used to determine the damage parameter, is computed by means of the spatial average of the local history variables in the vicinity of the point for which the stress is evaluated.
The peak load of the beam is very sensitive to the stress-strain response of the material directly above the notch.
For the stress-evaluation of this region, the areas below the notch are included in the averaging.
However, these areas are not strained during the loading process, so that the local history variable of the damage model will remain zero in these regions.
Therefore, the material above the notch is stronger than points further away from the notch along the ligament.
For $\mu \rightarrow 0.883$ the stress is limited by the local plasticity model.
On the other hand, for lower values of $\mu$, the plasticity model exhibits hardening.
Thus, a reduction of the unloading ratio results in an increase of the strength of the region above the notch and, consequently, in an increase of the peak load of the beam.
 
\section{Conclusion}

The present work deals with a damage-plasticity approach for the analysis of fracture in concrete.
Two approaches to obtain mesh-independent load-displacement curves were applied to a uniaxial bar subjected to tension and a three-point-bending test.
The first approach relies on an adjustment of the softening modulus with respect to the finite element size, whereas the second approach is based on nonlocal averaging of damage history variables.
The following conclusions can be drawn:
\begin{itemize}
\item For both approaches the present damage-plasticity model results in a mesh independent description of the softening and unloading regime of the load-displacement curves.
\item For the adjustment of the softening modulus, a change of the unloading stiffness does not influence the softening regime of the load-displacement curves for the finite element meshes used in the present study.
\item For the nonlocal model, the choice of the unloading stiffness influences strongly the load-displacement curves obtained for the three-point bending test.
This dependency is explained by the influence of the boundary near the notch, which affects the nonlocal averaging procedure.
\end{itemize}

The present results indicate that the nonlocal damage-plasticity model exhibits a strong boundary effect.
More research is required to develop boundary operators that overcome this influence of boundaries on nonlocal averaging.
This is one of the reasons why nonlocal modes are currently not yet implemented in commercial finite element codes.
Nevertheless, nonlocal models are more general than models based on the adjustment of the softening modulus with respect to the element size.
They are less sensitive to the alignment of the finite element mesh and produce even mesh-independent results, if localised and distributed failure modes coexist \cite{GraJir05}. 

\bibliographystyle{plainnat}
\bibliography{general}

\end{document}